\documentclass[preprint]{aastex}

\usepackage{graphics}

\shorttitle{Detection of [N~II] $\lambda$5755}
\shortauthors{Reynolds, Sterling, Haffner, /& Tufte}

\begin{document}

\title{Detection of [N ~II] $\lambda$5755 Emission from Low Density
       Ionized Interstellar Gas}

\author{R. J. Reynolds\altaffilmark{1}, N. C. Sterling\altaffilmark{1}, L.
   M. Haffner\altaffilmark{1}, and S. L. Tufte\altaffilmark{2}}

\altaffiltext{1}{Astronomy Department, University of Wisconsin,
    475 North Charter Street, Madison, WI 53706}
\altaffiltext{2}{Department of Physics, Lewis and Clark College, 0615 SW
    Palatine Hill Road, Portland, OR 97219; tufte@lclark.edu}

\begin{abstract}

The extremely faint, temperature sensitive ``auroral'' emission line
[N~II] $\lambda$5755 has been detected from the low density ionized gas
along the sight line toward $l = 130.^{\rm o}0$, b = $-7.^{\rm o}5$ using
the Wisconsin H$\alpha$ Mapper (WHAM).  The intensity of this emission
line, relative to the red nebular line [N ~II] $\lambda$6584, is found to
be twice that observed in bright, classical H~II regions surrounding O
stars.  This implies that the electron temperature of the ionized gas
along this sight line is about 2000 K higher than the H~II regions, and
that the enhanced [N ~II] $\lambda$6584/H$\alpha$ and [S ~II]
$\lambda$6716/H$\alpha$ intensity ratios in this low density gas are due
at least in part to an elevated temperature.

\end{abstract}


\section{Introduction}

Regions of nearly fully ionized hydrogen, having a temperature near
10$^4$ K and an electron density of order 10$^{-1}$ cm$^{-3}$, are wide
spread throughout the disk and halo of the Milky Way and other galaxies.
The source of the ionization and heating of this diffuse ionized gas is
not yet understood. However, important clues can be found in its optical
spectrum, which is characterized by high [N ~II] $\lambda$6584/H$\alpha$
and [S ~II] $\lambda$6716/H$\alpha$ line intensity ratios relative to
the ratios observed in the much brighter, higher density H~II regions
around O stars (e.g., Rand 1997; Haffner, Reynolds, and Tufte 1999).
Such elevated ratios indicate ionization/excitation conditions in the
diffuse ionized gas that differ significantly from conditions in the
classical H~II regions.

Although photoionization models incorporating a very dilute ionizing
radiation field (a low ionization parameter) are generally successful in
producing the elevated line ratios (e.g., Sokolowski 1991; Domg\"orgen \&
Mathis 1994; Greenawalt, Walterbos. \& Braun 1997), they fail to explain
some important details of the observations.  In particular, pure
photoionization models cannot account for the very high [N~II]/H$\alpha$,
[S~II]/H$\alpha$, and [O~II]/H$\alpha$ ratios (near unity and above)
observed far from the midplane, nor can they explain the observed
constancy of [S~II]/[N~II], even when the variations in [S~II]/H$\alpha$
and [N~II]/H$\alpha$ from direction to direction are large (e.g., see
Rand 1997, 1998; Ferguson et al 1996; T\"ullmann et al 2001).  Haffner et
al (1999) and Reynolds, Haffner, \& Tufte (1999, 2000) have shown that
these observations could be explained if the variations in
[S~II]/H$\alpha$ and [N~II]/H$\alpha$ were the result of variations in
the electron temperature of the emitting gas rather than variations in
the ionization parameter. Such temperature variations require the
existence of an additional source of heat in the low density gas that
dominates over the photoionization heating (Reynolds et al 1999).

One prediction of this idea is that the electron temperature in regions
of faint, low density ionized gas, where the [N~II]/H$\alpha$ ratios are
high, must be significantly greater than the temperature in the higher
density, bright H~II regions, where the [N~II]/H$\alpha$ ratio is
relatively low.  A standard method of determining electron temperatures
in ionized interstellar gas is to measure the
[N~II]$\lambda$5755/[N~II]$\lambda$6584 line intensity ratio.  Since
[N~II]$\lambda$6584 is one of the brightest lines emitted by diffuse
ionized gas, comparable in intensity to the H$\alpha$ (Haffner et al
1999), these nitrogen lines are a potentially useful diagnostic,
provided that a sufficiently sensitive spectrometer is available to
detect the much weaker 5755 \AA\ transition.  For densities below 10$^4$
cm$^{-3}$, the situation for the ionized regions considered here,
Osterbrock (1989) gives the volume emissivity ratio for the nitrogen
lines simply as

\begin{equation}
\frac{\epsilon (5755)}{\epsilon (6584)} = 0.192 \, e^{-25,000/T_e}.
\end{equation}\\
Below we present spectra of H$\alpha$, [N~II]$\lambda$6584, and
[N~II]$\lambda$5755 obtained with the Wisconsin H$\alpha$ Mapper (WHAM) in
a direction that samples faint low density gas and in directions
toward bright, higher density H~II regions. The results provide evidence
for a significant temperature difference between the two types of ionized
regions.

\section{Observations and Results}

The observations were made with the WHAM spectrometer, located on Kitt
Peak and operated remotely from Madison, Wisconsin (Haffner et al 1999;
Tufte 1997).  In 1998 October and 1999 November long integration spectra
were obtained of a 3.8 \AA\ (200 km s$^{-1}$) wide spectral interval
centered near the Local Standard of Rest (LSR) velocity of the
[N~II]$\lambda$5755 line.  The spectral resolution was 12 km s$^{-1}$ and
the beam on the sky was $1.^{\rm o}0$ in diameter.  The observations
consisted of a series of integrations (each 300 s or 600 s) toward $l =
130.^{\rm o}0$, b = $-7.^{\rm o}5$, which samples diffuse low density
gas. Between each of these ``on-source'' observations, an integration of
approximately equal length was made toward one of a number of higher
Galactic latitude, even fainter ``off-source'' directions: $135.^{\rm
o}0, -30.^{\rm o}0$; $135.^{\rm o}0, -35.^{\rm o}0$; $135.^{\rm o}0,
-40.^{\rm o}0$; $142.^{\rm o}5, -22.^{\rm o}5$; and $157.^{\rm o}0,
-24.^{\rm o}5$.  This technique is described by Tufte, Reynolds, \&
Haffner (1998) and has been used to detect very faint optical emission
lines from high velocity H~I clouds.  Observations of the fainter,
``off-source'' directions are necessary because the spectra are
contaminated by very weak, unidentified atmospheric emission lines that
are comparable to or brighter than the interstellar emission (Hausen et
al 2000; Tufte et al 1998).  By subtracting the average of the
``off-source'' spectra from the ``on-source'' spectrum, we obtain a pure
interstellar spectrum of [N~II]$\lambda$5755 toward $l = 130.^{\rm o}0$,
b = $-7.^{\rm o}5$.  Because diffuse ionized gas covers the entire sky,
this process also subtracts a small amount of the interstellar emission.
For example, the average H$\alpha$ and [N~II]$\lambda$6584 intensity in
the ``off-source'' directions is 16\% and 20\%, respectively, that toward
$130.^{\rm o}0$, $-7.^{\rm o}5$.  For consistency, the same procedure was
used to obtain spectra of H$\alpha$ and [N~II]$\lambda$6584 for this
direction, as well as H$\alpha$, [N~II]$\lambda$6584, and
[N~II]$\lambda$5755 for seven bright H~II regions around O stars (this
``on'' minus ``off'' technique was not really necessary in these cases
because the interstellar emission was always much brighter than the
atmospheric emission ---except the geocoronal H$\alpha$ line, which is
removed by a different procedure; see Reynolds et al 1998).

The direction $130.^{\rm o}0$, $-7.^{\rm o}5$ is within the region mapped
in H$\alpha$, [N~II]$\lambda$6584, and [S~II]$\lambda$6716 by Haffner et
al (1999).  This particular sight line was selected because the H$\alpha$
emission is relatively bright, making it possible to detect the
exceedingly faint [N~II]$\lambda$5755 line, and because the
[N~II]$\lambda$6584/H$\alpha$ and [S~II]$\lambda$6716/H$\alpha$ ratios
(0.43 and 0.27, respectively) are characteristic of diffuse low density
gas and are significantly higher than the ratios in the bright H~II
regions (e.g., for the H~II regions considered here, the average ratios
are 0.27 and 0.08, respectively). Radial velocity components ranging
from near the local standard of rest (LSR) out to about $-80$ km
s$^{-1}$ are clearly present in the H$\alpha$ and [N~II]$\lambda$6584
profiles, indicating that the ionized gas is spread along a large portion
of the sight line.  A very extended, faint H~II region associated with
the B1 star $\phi$ Per (4$^{\rm o}$ away and 220 pc distant) appears to
dominate the profile near 0 km s$^{-1}$, while the emission at the most
negative velocities is associated with gas more than 2 kpc distant and 
300 pc from the midplane in the Perseus arm or beyond (see Haffner et al 
1999).

The resulting ``on'' minus ``off'' spectra are shown in Figure 1 for
$130.^{\rm o}0$, $-7.^{\rm o}5$ and for one of the H~II regions.  The
[N~II]$\lambda$5755 spectrum toward $130.^{\rm o}0$, $-7.^{\rm o}5$ has a
total ``on-source'' integration time of 12,300 s (19,500 s ``off-source'')
and shows the first detection of this nitrogen line from low density
ionized gas.  Integration times for H$\alpha$ and [N~II]$\lambda$6584
in this direction were 660 s and 1020 s (720 s and 1200 s ``off''),
respectively, while for the H~II regions the integration times ranged from
30 s to 60 s for the H$\alpha$, 60 s to 120 s for [N~II]$\lambda$6584, and
120 s to 1200 s for [N~II]$\lambda$5755.

Line intensities were calculated by fitting gaussian emission components
to the spectra.  For $130.^{\rm o}0$, $-7.^{\rm o}5$ the
[N~II]$\lambda$6584 profile indicates four radial velocity components at
$-81$ km s$^{-1}$, $-61$ km s$^{-1}$, $-40$ km s$^{-1}$, and $-7$ km
s$^{-1}$.  In the H$\alpha$ spectrum these components are not as well
defined because of the larger thermal broadening of the hydrogen.
Gaussians at these four velocities were fitted to each of the three
spectra towards $130.^{\rm o}0$, $-7.^{\rm o}5$, while single gaussians
were used for the H~II region spectra.  This produced good fits in all
cases, and the results are listed in Table 1.  The intensity
calibrations of the nitrogen lines are based on the H$\alpha$
calibration with small corrections for instrument response and
atmospheric transmission.  Since the principal goal is to compare
emission line ratios between the $130.^{\rm o}0$, $-7.^{\rm o}5$
direction and the bright H~II regions, small systematic errors in the
absolute calibrations of the individual lines will not affect our
conclusions.  For [N~II]$\lambda$5755, the position of the baseline was
the principal source of uncertainty for the component intensities, and
the listed uncertainties were derived by examining many fits using
various adopted baselines.

Because interstellar extinction has only a small effect on the emission
line ratios, no extinction corrections have been made.  The exciting stars
for the bright H~II regions in our sample have values for E$_{B-V}$ that
range from 0.12 to 0.71 (see Reynolds 1988), which according to Mathis
(1983) would require correction factors of 1.03 to 1.10 to the observed
[N~II]$\lambda$5755/[N~II]$\lambda$6584 intensity ratios.  The H~I column
density toward $130.^{\rm o}0$, $-7.^{\rm o}5$ is 1.5 $\times 10^{20}$
cm$^{-2}$ (Hartmann and Burton 1998), impling an E$_{B-V} \approx 0.3$,
similar to the average (0.4) toward the H~II regions.  Thus the comparison
of the ratios between this sight line and the H~II regions should not be
affected significantly by extinction.  The principal effect of applying
extinction corrections would be to increase all the temperatures derived
from equation (1) by about 200~K.

The [N~II]$\lambda$5755/[N~II]$\lambda$6584 and
[N~II]$\lambda$6584/H$\alpha$ line intensity ratios for the H~II regions
and for $130^{\rm o}.0$, $-7.^{\rm o}5$ are plotted in Figure 2.  The
results listed in Table 1 for the individual velocity components toward
$130^{\rm o}.0$, $-7.^{\rm o}5$ suggest that there may be large
variations in the ratios from component to component.  Components 1 and
4, for example, appear to have higher
[N~II]$\lambda$5755/[N~II]$\lambda$6584 ratios than Components 2 and 3.
However, because of severe blending between the components, except for
component 4, and because of the relatively low signal to noise of the
[N~II]$\lambda$5755 spectrum, only the results for component 4 and for
the integrated line profile (the sum of all four velocity components)
are plotted in Figure 2.

\section{Discussion and Conclusions}

The emission line [N~II]$\lambda$5755 has been detected for the first
time in the faint ``diffuse background'', making it possible to explore
directly the electron temperature in the low density ionized gas.
Figure 2 clearly shows that toward $130.^{\rm o}0$, $-7.^{\rm o}5$ not
only does the gas have a larger [N~II]$\lambda$6584/H$\alpha$ ratio than
the bright H~II regions (a long recognized characteristic of diffuse
ionized gas), but that the [N~II]$\lambda$5755/[N~II]$\lambda$6584 ratio
is also larger, implying that the faint, diffuse gas has a significantly
higher electron temperature T$_e$ than the gas in the H~II regions.
For the H~II regions, derived values of T$_e$ from equation (1) range
from 6400 K to 6900 K, while toward $130.^{\rm o}0$, $-7.^{\rm o}5$,
T$_e$ = 9400 $\pm$ 400 K (component 4) and 8400 $\pm$ 500 K (the
integrated profile).  Thus, at least for this sight line the enhanced
[N~II]$\lambda$6584/H$\alpha$ is due in part to an elevated temperature.

To examine more quantitatively the relationship between the intensity
ratios and electron temperature, we have plotted as a solid line in
Figure 2 the ratios predicted for various temperatures based upon
equation (1) and the relationship between [N~II]$\lambda$6584/H$\alpha$
and temperature presented by Haffner et al (1999).  Interestingly, nearly
all of the observations lie above this line. This suggests that no single
temperature can account for the observed ratios, not even toward the
bright H~II regions. The relationship between T$_e$ and [N~II]/H$\alpha$
is based on the fact that in photoionized interstellar gas, N$^+$/N
$\approx$ H$^+$/H (e.g., Sembach et al 2000).  Thus the offset between
the solid line and the observations could be explained if there were a
large amount of N$^{++}$, specifically, if N$^{++}$ $\approx$ N$^{+}$.  
While this may be possible for some of the H~II regions, it cannot be the
explanation for the $130.^{\rm o}0$, $-7.^{\rm o}5$ observations, since
the high [S~II]/H$\alpha$ ratio implies a relatively low ionization state
for the gas (Haffner et al 1999). A large error in the adopted value for
the gas phase abundance N/H also seems improbable (see Meyer, Cardelli,
\& Sofia 1997 and Afflerback, Churchwell, \& Werner 1997).  A more likely
explanation is that a range of temperatures is present along the line of
sight.  In this case [N~II]$\lambda$5755, with a higher excitation energy
(4.04 eV) than that (1.89 eV) of [N~II]$\lambda$6584, is produced
preferentially in the higher temperature regions, resulting in a
temperature derived from [N~II]$\lambda$5755/[N~II]$\lambda$6584 that is
higher than the temperature derived from [N~II]$\lambda$6584/H$\alpha$.
The dashed curve in Figure 2 represents a very simple case in which there
is an equal mixture (by emission measure) of gas at two temperatures,
half at 5000 K and half at the higher temperature indicated along the
curve.  While this fit is certainly not unique, it demonstrates that a
mixture of temperatures can readily explain the observed ratios, and that
toward $130.^{\rm o}0$, $-7.^{\rm o}5$ at least some of the gas has a
significantly higher temperature than in the denser H~II regions.

The direction $130.^{\rm o}0$, $-7.^{\rm o}5$ was chosen in part because
of its relatively high H$\alpha$ intensity.  If an inverse relationship
exists between temperature and density, then sight lines with fainter
emission (lower densities) may have an even higher temperature than that
derived for $130.^{\rm o}0$, $-7.^{\rm o}5$.  Furthermore, the derived
temperature for $130.^{\rm o}0$, $-7.^{\rm o}5$ itself would be an
underestimate in this case, since higher temperature ``offs'' would
result in a slight oversubtraction of [N~II]$\lambda$5755 relative to
[N~II]$\lambda$6584.

The detection of the [N~II] $\lambda$5755 emission has therefore provided
strong evidence for a temperature difference between diffuse, low density
ionized gas and denser, classical H~II regions.  However, to explore more
fully temperatures and possible variations in temperature within the
diffuse ionized medium itself, many additional observations will be
needed, observations in directions farther from the midplane, where the
[N~II]$\lambda$6584/H$\alpha$ and [S~II]$\lambda$6717/H$\alpha$ ratios
are even larger (and temperatures are higher?) and where the influence of
individual O and B stars is less.  Unfortunately, because
[N~II]$\lambda$5755 is so weak, the use of this line for extensive
studies of such faint regions is problematic; in fact, all single-ion
diagnostics, which are insensitive to abundance and ionization state,
appear to involve lines that are extremely faint.  Less ideal, but
brighter, temperature sensitive line ratios, such as [O~II]
$\lambda$3727/H$\alpha$, need to be investigated (Ferguson et al 1996;
Otte et al 2001). Since [O~II] has an excitation energy of 3.3 eV and an
intensity comparable to that of H$\alpha$, [O~II]/H$\alpha$ perhaps could
be used as a temperature diagnostic for diffuse ionized gas, particularly
in combination with observations of [O~I] $\lambda$6300 and [O~III]
$\lambda$5007 to monitor oxygen's ionization state.

\acknowledgments

This work was supported by the National Science Foundation through grant
AST 96-19424.

\clearpage

\begin{deluxetable}{cccc}
\tablecolumns{4}
\tablewidth{0pc} 
\tablecaption{Results}
\tablehead{
\colhead{Name of} & \colhead{H$\alpha$} & 
\colhead{$\frac{[NII]6584}{H\alpha}$} & 
\colhead{100 $\times \frac{[NII]5755}{[NII]6584}$} \\
\colhead{Region\tablenotemark{a}} & \colhead{(R)\tablenotemark{b}} 
& \colhead{(energy)\tablenotemark{c}} & \colhead{(energy)\tablenotemark{c}}
}
\startdata 
C1   & 1.51 $\pm$ 0.08 & 0.38 $\pm$ 0.06 & 1.3 $\pm$ 0.6 \\
C2   & 2.14 $\pm$ 0.10 & 0.21 $\pm$ 0.03 & 0.3 $\pm$ 0.3 \\
C3   & 2.06 $\pm$ 0.10 & 0.59 $\pm$ 0.04 & 0.4 $\pm$ 0.2 \\
C4   & 4.15 $\pm$ 0.14 & 0.52 $\pm$ 0.04 & 1.33 $\pm$ 0.12 \\
C(Total)   & 10.11 $\pm$ 0.18 & 0.43 $\pm$ 0.01 & 0.96 $\pm$ 0.21 \\
S117 & 800\tablenotemark{d} & 0.202 $\pm$ 0.005 & 0.45 $\pm$ 0.03 \\
S132 & 136 $\pm$ 2\tablenotemark{e} & 0.218 $\pm$ 0.005 & 0.41 $\pm$ 0.03 \\
S184 & 145 $\pm$ 2\tablenotemark{e} & 0.294 $\pm$ 0.008 & 0.44 $\pm$ 0.03 \\
S220 & 339 $\pm$ 5 & 0.383 $\pm$ 0.012 & 0.38 $\pm$ 0.02 \\
S252 & 202 $\pm$ 4\tablenotemark{e} & 0.270 $\pm$ 0.009 & 0.49 $\pm$ 0.03 \\
S261 & 68 $\pm$ 1\tablenotemark{e} & 0.353 $\pm$ 0.009 & 0.50 $\pm$ 0.06 \\
S264 & 157 $\pm$ 2 & 0.216 $\pm$ 0.006 & 0.42 $\pm$ 0.04 \\ 
\enddata
\tablenotetext{a}{C1, C2, C3, C4 refer to components at radial velocities -81, -61, -40, and -7 km s$^{-1}$, respectively, for the line of sight $130^{\rm o}, -7.5^{\rm o}$; C(Total) is the velocity integrated line profile.  The ''S'' designation for the HII regions refers to the Sharpless Catalog number (Sharpless 1959).}
\tablenotetext{b}{1R = $10^{6}/4\pi$ photons cm$^{-2}$ s$^{-1}$ sr$^{-1}$, or $2.41 \times 10^{-7}$ erg cm$^{-2}$ s$^{-1}$ sr$^{-1}$ for H$\alpha$.}
\tablenotetext{c}{These are intensity ratios (not photon ratios).}
\tablenotetext{d}{The H$\alpha$ intensity of this region (NGC 7000) is used as the absolute intensity standard (Scherb 1981).}
\tablenotetext{e}{These regions do not fill WHAM's $1^{\rm o}$ diameter beam (listed H$\alpha$ intensities are averaged within the beam).}
\end{deluxetable}

\clearpage

\begin{figure}[htbp]
  \begin{center}
    \includegraphics{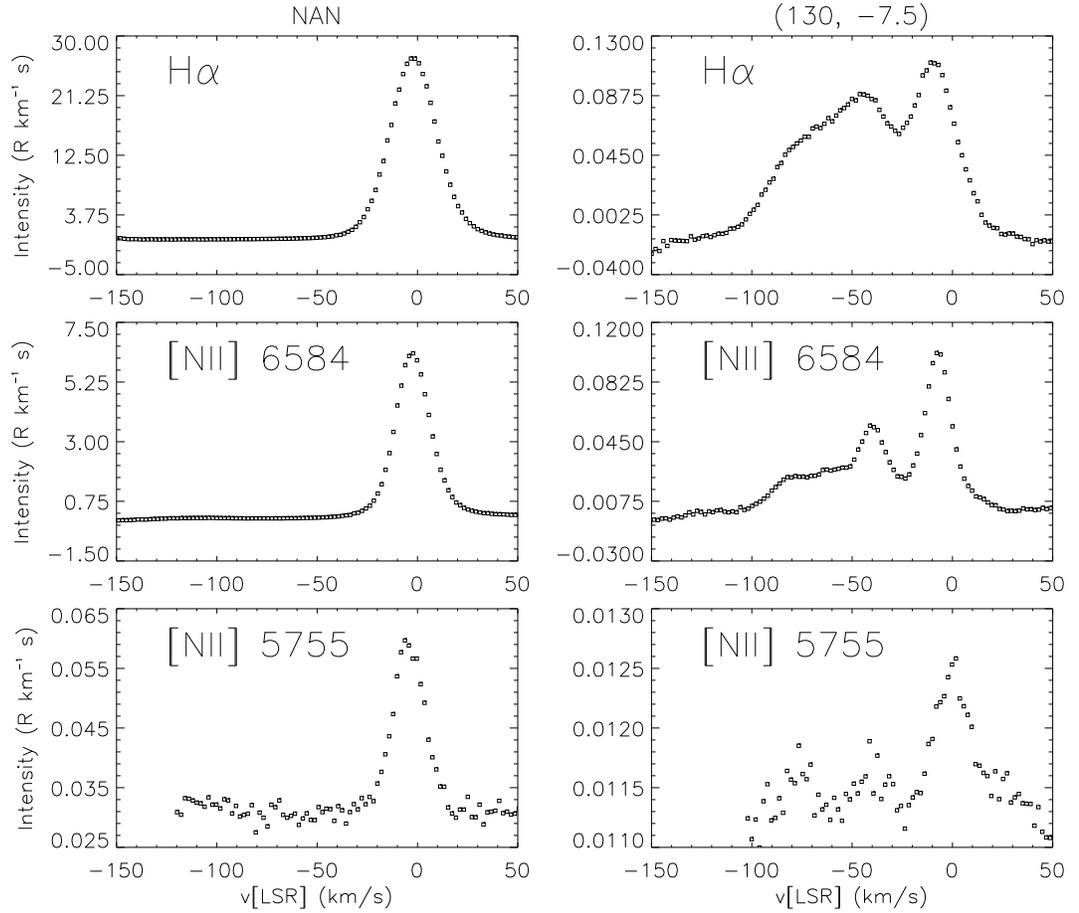}
    \caption{The resulting WHAM spectra of H$\alpha$, [N~II]
      $\lambda$6584, and [N~II] $\lambda$5755 toward $l = 130.^{\rm
        o}0$, b = $-7.^{\rm o}5$ and toward the O star H~II region
      S117 (North American Nebula).  The vertical scatter in the data
      points provides a measure of the observational uncertainty.
      Except for the $\lambda5755$ spectra, the uncertainties are
      comparable to or less than the size of the symbol.}
  \end{center}
\end{figure}

\begin{figure}[htbp]
  \begin{center}
    \includegraphics{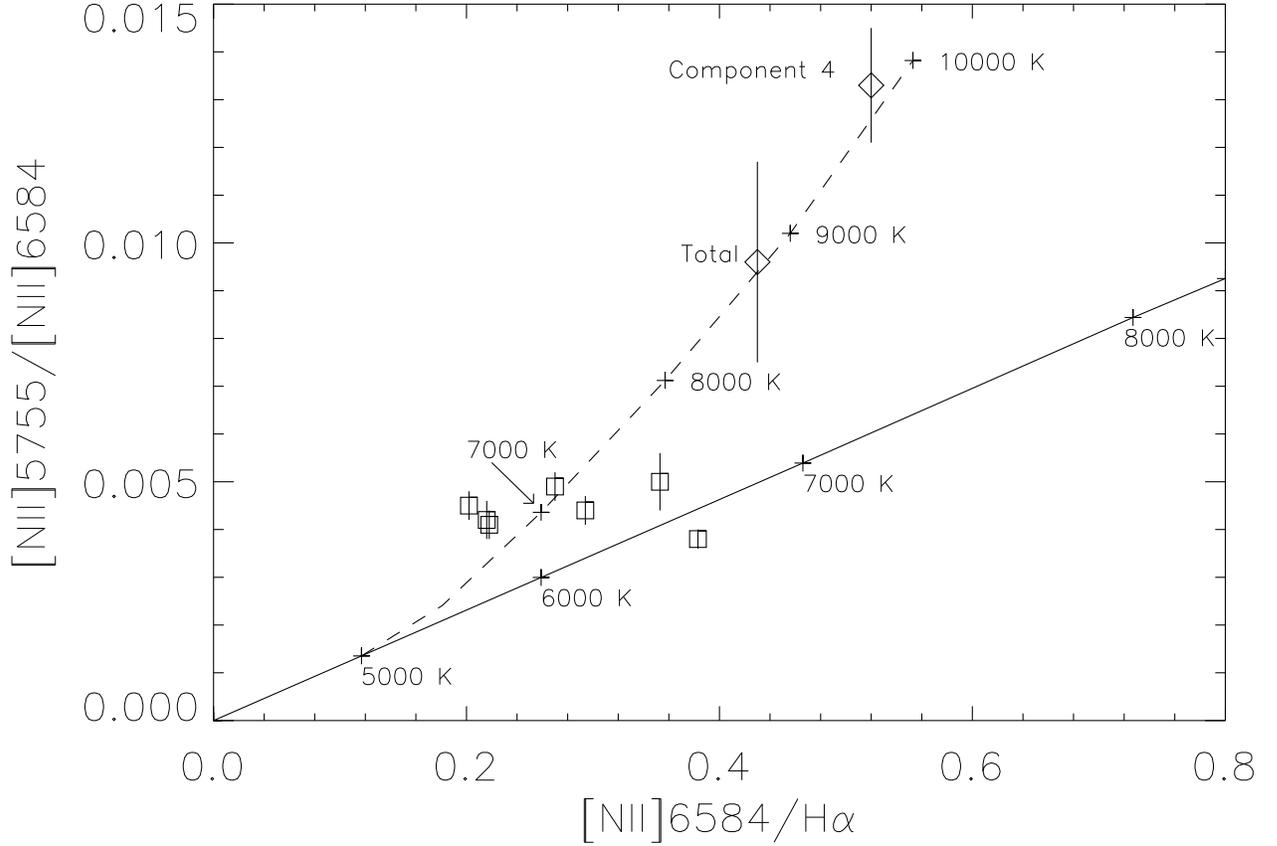}
    \caption{The 5755/6584 line intensity ratio plotted vs
      6584/H$\alpha$ for $130.^{\rm o}0$, $-7.^{\rm o}5$ ($\Diamond$),
      and for the seven bright H~II regions ($\Box$).  The solid line
      denotes the expected ratio based upon Osterbrock (1989) and
      Haffner et al (1999) for an emission region of uniform electron
      temperature.  The dashed curve is the expected relationship if
      the emission is from an equal mixture (by emission measure) of
      gas at 5000 K and gas at a higher temperature indicated along
      the curve (see text).}
  \end{center}
\end{figure}

\end{document}